\documentclass[12pt]{article}
\usepackage{graphicx}

\begin{document}

\centerline{\bf \large Absence of ferromagnetism in Ising model}
\bigskip

\centerline{\bf \large  on directed Barabasi-Albert network}

\bigskip
\noindent
Muneer A. Sumour, Physics Department, Al-Aqsa University,  P.O.B. 4051, Gaza,  
Gaza Strip, Palestinian Authority

\noindent
M.M. Shabat, Physics Department, Islamic University, P.O.B. 108, Gaza,
Gaza Strip, Palestinian Authority

\noindent
D. Stauffer, Institute for Theoretical Physics, Cologne University, 
D-50923 K\"oln, Euroland

Email:  

\noindent
Shabat@mail.iugaza.edu, msumoor@yahoo.com, stauffer@thp.uni-koeln.de

\bigskip
Abstract:

With up to 7 million spins, the existence of spontaneous magnetization of 
Ising spins  on directed Barabasi-Albert networks is investigated by 
Monte Carlo simulations. We confirm our earlier result that the magnetization 
for different temperatures $T$ decays after a characteristic time $\tau(T)$, 
which we extrapolate to diverge at zero temperature by a modified Arrhenius law,
or perhaps a power law. 

\bigskip

\begin{figure}[hbt]
\begin{center}
\includegraphics[angle=-90,scale=0.5]{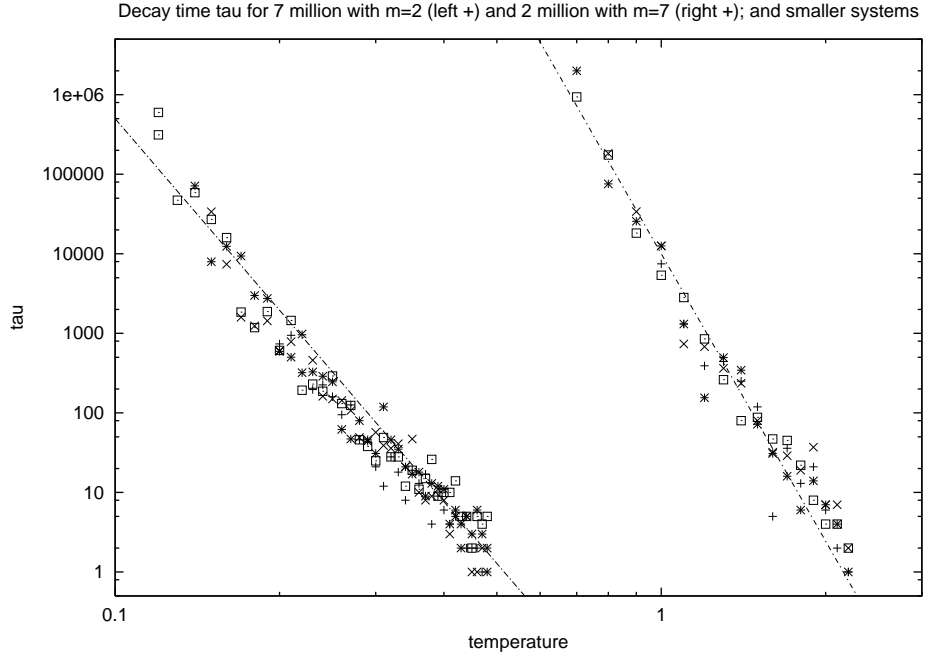}
\end{center}
\caption{
Characteristic time for $M(\tau) = 3/4$ using 7 million spins for $m=2$ 
neighbours and 7 million spins for $m=7$ neighbours (+). Ten, hundred, and
thousand times smaller systems are denoted by x, stars, and squares. We plot
the median over nine samples in this log-log plot. The two straight lines
have negative slopes 8 (left) and 12 (right). 
}
\end{figure}

\begin{figure}[hbt]
\begin{center}
\includegraphics[angle=-90,scale=0.50]{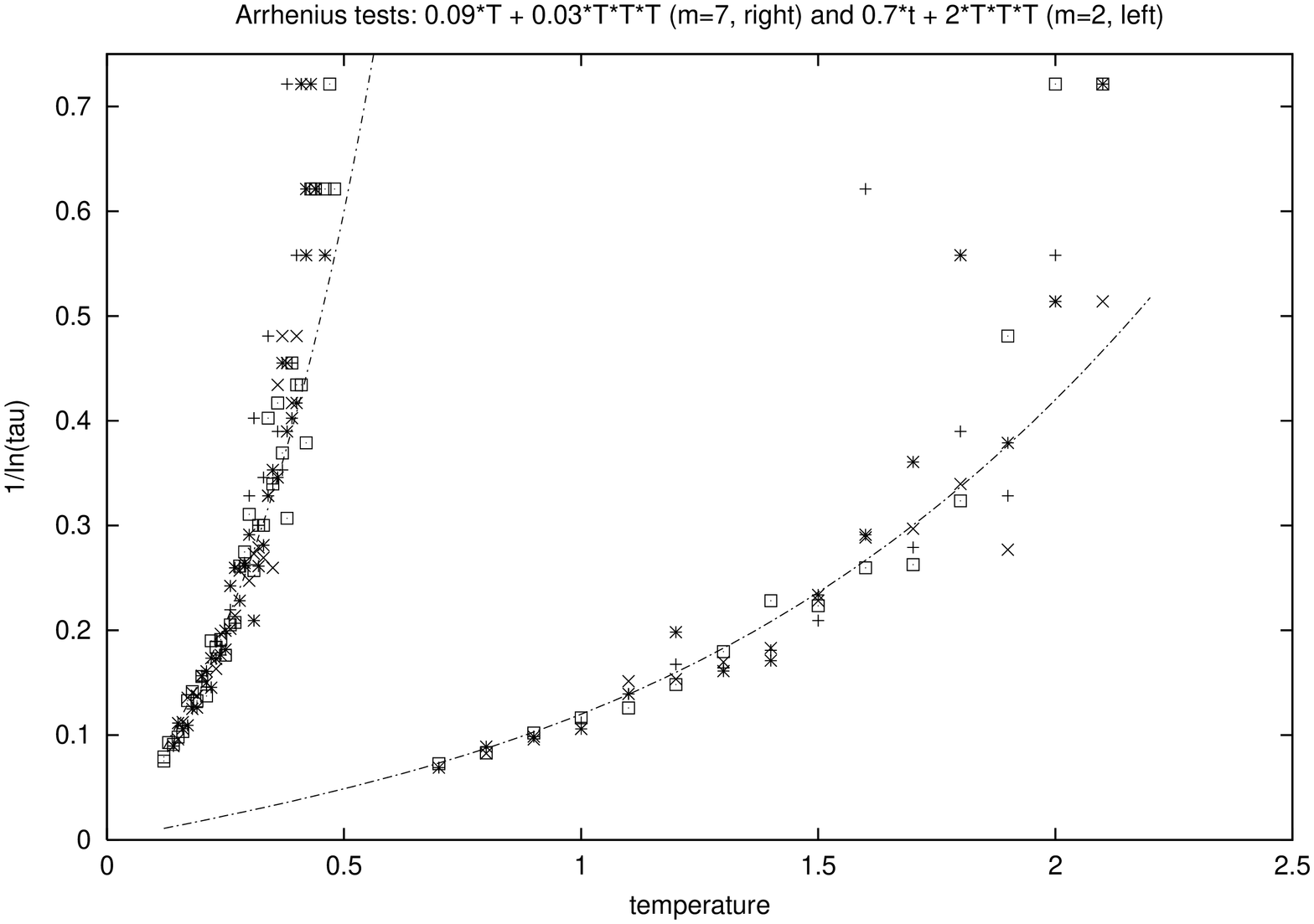}
\end{center}
\caption{
Same data (except for short times) plotted as 1/ln($\tau$) versus T.}
\end{figure}

{\bf Introduction}: 

The Ising magnet is since decades a standard tool of computational physics [1].
We apply it here to scale-free networks [2], where previous simulations  [3]
indicated a Curie temperature increasing logarithmically with increasing system
size $N$. In contrast to that work we use here directed [4] as opposed to 
undirected networks and then apply the standard Glauber kinetic Ising model [1]
to the fixed network. We try to improve our previous note [5] by making the 
system an order of magnitude larger, correcting a programming error, and 
comparing networks with two and seven neighbours.

\bigskip
{\bf Directed scale-free network}:

Putting Ising spins onto the sites (vertices, nodes) of a network, we  
simulate our Ising magnetic model on directed Barabasi-Albert networks. 
The Barabasi-Albert network is grown such that the probability of a new site
to be connected to one of the already existing sites is proportional to the 
number of previous connections to this already existing site: The rich get 
richer. In this way each new site selects exactly $m$ old sites as neighbours.

Then each spin is influenced by the fixed number $m$ of neighbours which it had 
selected when joining the network. It is not influenced by other spins which 
selected it as neighbour after it joined the network.

At each step, a new spin is added which builds $m$ new connections {\tt neighb},
randomly to already existing spins. The probability for an existing spin to be 
chosen as neighbour is proportional to the number of its neighbours. 

\bigskip
{\bf Ising simulations}:

The Ising interaction energy is
$$ E = -J \sum_i \sum_k S_i S_k \quad (S_i = \pm 1) $$
where the sum over $k$ goes only over the $m$ neighbours which site $i$ had 
selected when it joined the network.  We measure the temperature in units of 
the usual Curie temperature of the square-lattice Ising model. 

First we initialize a directed Barabasi-Albert network of $N$ sites
with $m$ neighbours (all 
$m$ initial spins are connected with each other and themselves), here $m=7$. 
We put an Ising spins onto every site, with all spins up, because we test here 
for ferromagnetism. Then with the standard Glauber (heat bath) Monte Carlo 
algorithm spins we search for thermal equilibrium at positive temperature. 
Time $t$ is measured in Monte Carlo steps MCS per spin.

As shown earlier [5] the magnetization $M(t)$ as a function of time $t$ shows
strong fluctuations and sometimes very rapid changes in a very short interval.
These problems became worse when we corrected a programming error. Now the
magnetization reduces to a temperature-dependent metastable value after the
first time step, stays there for a long time, and then flips or tries to flip.
We follow 
[5] and define a characteristic time $\tau$ as that time where the magnetization
has decreased to 3/4 of its initial value, Fig.1. This double-logarithmic plot
first suggests a power-law divergence $\tau(T \rightarrow 0) \rightarrow 
\infty$. But we see slight curvature, unusually high exponents near 10, and
an exponent varying with the number $m = 2$ or 7 of neighbours. More plausible
thus is the modified Arrhenius plot of Fig.2 which suggests $\tau \propto
\exp({\rm const}/T)$ for low temperatures, or 1/ln$(\tau) \propto T + \dots$ 
This const in the exponential is about 1.4 for $m=2$ and about 11 for $m=7$.

\bigskip
{\bf Conclusion}:

In this way we confirmed the asymptotic Arrhenius extrapolation $1/\ln \tau 
\propto T$ of [5], meaning that at all finite temperatures the 
magnetization eventually vanishes: No ferromagnetism.

\bigskip
{\bf References}:

\parindent 0pt

[1] David P. Landau, Kurt Binder, A guide to Monte Carlo simulation in 
statistical physics; Cambridge University Press (2002).

[2] R. Albert and A.L. Barabasi, Rev. Mod. Phys. 74, 47 (2002).

[3] A. Aleksiejuk, J. A. Holyst, D. Stauffer, Physica A 310, 260 (2002);
J.O. Indekeu,  Physica A 333, 461 (2004).

[4] D. Stauffer and H. Meyer-Ortmanns, Int. J. Mod. Phys.C 15, 241 (2004).

[5] M.A. Sumour and M.M. Shabat, Int. J. Mod. Phys. C 16, no. 4 (2005) =
e-print cond-mat/0411055 at www.arXiv.org
\end{document}